\documentclass[9pt, twocolumn]{article}
\usepackage{geometry}                
\usepackage{graphicx}
\usepackage{authblk}
\usepackage{amssymb,amsmath}
\usepackage{epstopdf}
\usepackage{enumerate}
\usepackage{url}
\usepackage{sectsty}
\usepackage{subcaption}

\sectionfont{\large}

\title{Comment on ``Distinguishing Classical and Quantum Models\\ for the {D-Wave} Device''}
\author[*]{Seung Woo Shin}
\author[$\dag$]{Graeme Smith}
\author[$\dag$]{John A.~Smolin}
\author[*]{Umesh Vazirani}
\affil[*]{\textit{Computer Science division, UC Berkeley, USA.}}
\affil[$\dag$]{\textit{IBM T.J.~Watson Research Center, Yorktown Heights, NY 10598, USA.}}
\date{}                                           

\begin{document}
\maketitle

{\textbf{\textit{ The SSSV model~\cite{SSSV} is a simple classical model that achieves excellent correlation with published experimental data on the D-Wave machine's behavior on random instances of its native problem~\cite{Boixo1}, thus raising questions about how ``quantum'' the D-Wave machine is at large scales. In response, a recent preprint by Vinci et al.~\cite{Vinci} proposes a particular set of instances on which the D-Wave machine behaves differently from the SSSV model. In this short note, we explain how a simple modeling of systematic errors in the machine allows the SSSV model to reproduce the behavior reported in the experiments of \cite{Vinci}.}}}

In the SSSV model~\cite{SSSV} for the D-Wave machine, qubits are modeled as classical magnets coupled through nearest-neighbor Coulomb interaction and subject to an external magnetic field. Moreover, the finite temperature of the device is modeled by performing a Metropolis update at each step. The results in~\cite{SSSV} showed that the model shows excellent correlation with published data about the input-output behavior of the D-Wave machine on randomly chosen input instances~\cite{Boixo1}. Nevertheless is it possible that there are other classes of input instances on which the D-Wave machine exhibits ``truly quantum" behavior? This is a question of central importance in the evaluation of the D-Wave architecture.

An affirmative answer requires exhibiting a regime in which classical models such as SSSV fail to reproduce the behavior of the D-Wave machine. Of course the SSSV model is extremely rudimentary, and was not meant to be an exact model for the D-Wave machine. For example, it makes no attempt to model details of the D-Wave machine such as errors in control of external fields and interaction strengths. So any such exhibited regime must either be sufficiently robust so that it can be argued that detailed modeling of the machine is unnecessary, or it must differentiate the behavior of D-Wave from reasonable elaborations of the SSSV model. Of course for the regime to be meaningful, there should also be a plausible computational benefit to the phenomenon in question. 

\begin{figure}[!t]
\centering
\includegraphics[width=0.5\columnwidth]{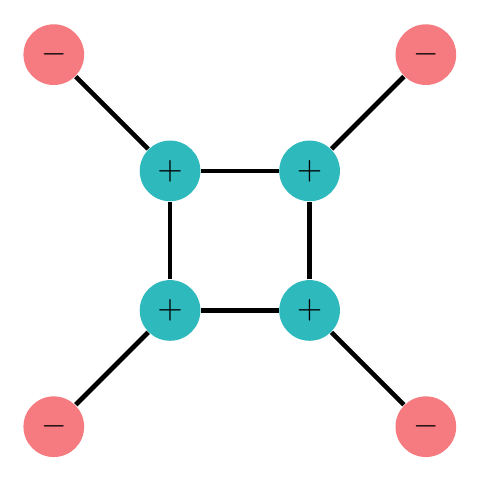}
\caption{The problem Hamiltonian used in the experiments of \cite{Vinci}. All couplings are ferromagnetic, whereas there is a local $z$-field applied in the $+$ direction for the four ``core'' spins, and in the $-$ direction for the four ``peripheral'' spins. Formally, the Hamiltonian is defined as $H = -\sum_i h_i \sigma_i^z -\sum_{i<j} J_{ij} \sigma_i^z\sigma_j^z$. The local field $h_i$ is set to be $1$ if $i$ is a core spin, and $-1$ otherwise. The coupling strength $J_{ij}=1$ for every edge $i\sim j$. Figure is from \cite{Vinci}.}\label{problem}
\end{figure}
\begin{figure}[!t]
\centering
\includegraphics[width=0.95\columnwidth]{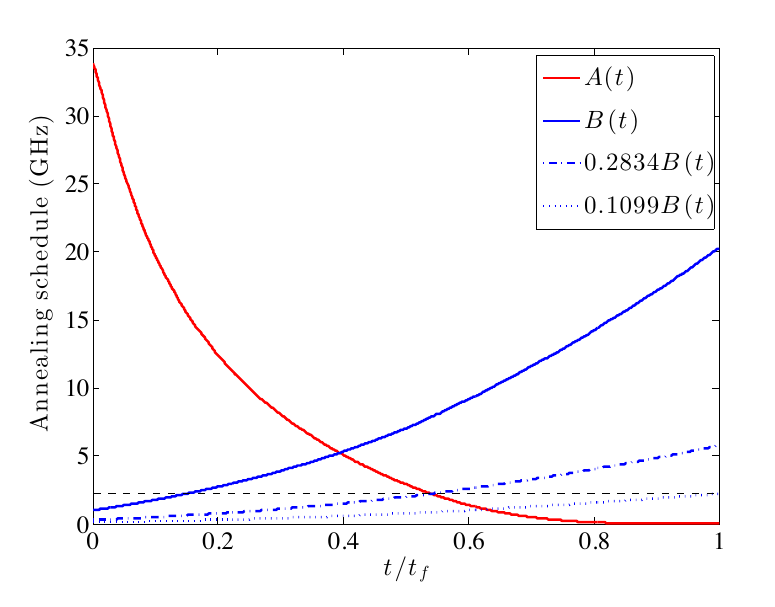}
\caption{The solid curves represent the annealing schedule of D-Wave Two. Dotted blue curves represent the effective annealing schedule for cases $\alpha =0.2834$ and $\alpha=0.1099$. The dotted black line represents the system temperature. Figure is from \cite{Vinci}.}\label{Vinci_schedule}
\end{figure}

A recent preprint by Vinci et al.~\cite{Vinci} reports that the behavior of the D-Wave machine and the SSSV model differ on a particular set of instances. Fig.~\ref{problem} depicts the problem Hamiltonian used in the experiments of \cite{Vinci}. The Hamiltonian has a 17-fold degenerate ground space with ground energy $-8$. It is easy to see that all sixteen states with the four core spins pointing in the $+$ direction are ground states. There is one more ground state in which all eight spins point in the $-$ direction. \cite{Vinci} call the first sixteen ground states the ``clustered'' ground states because they are connected by local spin flips of the peripheral spins, whereas they refer to the last ground state as the ``isolated'' ground state.

\begin{figure*}[!t]
\centering
\includegraphics[width=0.95\textwidth]{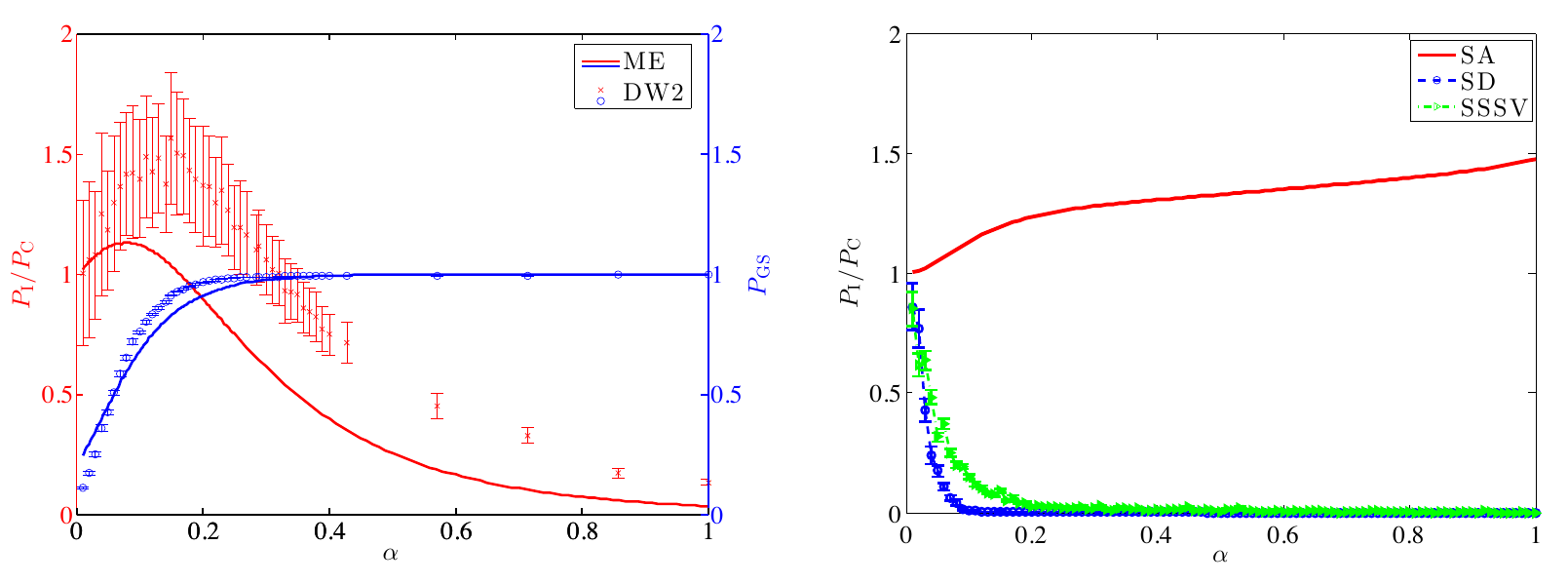}
\caption{Experimental and numerical results from \cite{Vinci}. DW2, ME, SA, SD, and SSSV represent D-Wave Two, quantum adiabatic master equation, simulated anneailng, Smolin-Smith model \cite{Smolin}, and SSSV model respectively. $P_{GS}$ indicates the probability of finding one of the seventeen ground states.}\label{Vinci_results}
\end{figure*}

This Hamiltonian was previously used in \cite{8qubit} to distinguish the behavior of the D-Wave machine from that of simulated annealing. As the problem size is fairly small, the D-Wave machine almost always succeeds in finding one of the 17 ground states, as did simulated annealing. To distinguish between the two, \cite{8qubit} considered the quantity $P_I/P_C$, where $P_I$ is the probability of seeing the isolated ground state at the end of the process and $P_C$ is the probability of seeing a clustered ground state divided by $16$. Experiments revealed that the D-Wave machine and the adiabatic Markovian master equation preferred the clustered ground state ($P_I/P_C<1$), whereas simulated annealing preferred the isolated ground state ($P_I/P_C>1$). A simple experiment confirms that the SSSV model also agrees with the behavior of D-Wave and the master equation ($P_I/P_C<1$).

To distinguish between D-Wave and SSSV, Vinci et al.~\cite{Vinci} perform a more elaborate version of this experiment with an additional control variable $\alpha$ which represents the scale of the final Hamiltonian. Namely, the machine is programmed to implement the time-dependent Hamiltonian $H(t)=A(t)H_0 +\alpha B(t)H_f$, where $\alpha$ is varied in the range $[0,1]$ as in Fig.~\ref{Vinci_schedule}. As shown in Fig.~\ref{Vinci_results}, they find that the machine and the adiabatic quantum master equation prefer the isolated ground state ($P_I/P_C>1$) when $\alpha$ is small, whereas the SSSV model always prefers the clustered ground state ($P_I/P_C<1$) at all values of $\alpha$.

It is illuminating to examine more closely the small $\alpha$ regime, where D-Wave and SSSV differ. Since in this regime the coupling strength is very small, this may be thought of as the ``classical regime'' where the machine is expected to be driven mostly by thermal noise rather than quantum effects. Moreover, as can be seen in Fig.~\ref{Vinci_schedule}, when $\alpha$ is small, it is only after the transverse field $A(t)$ has almost completely died out that the problem Hamiltonian becomes strong enough to be able to overcome the system temperature,  therefore effectively making the annealing schedule trivial. 

\begin{figure}[!t]
\centering
\includegraphics[width=0.95\columnwidth]{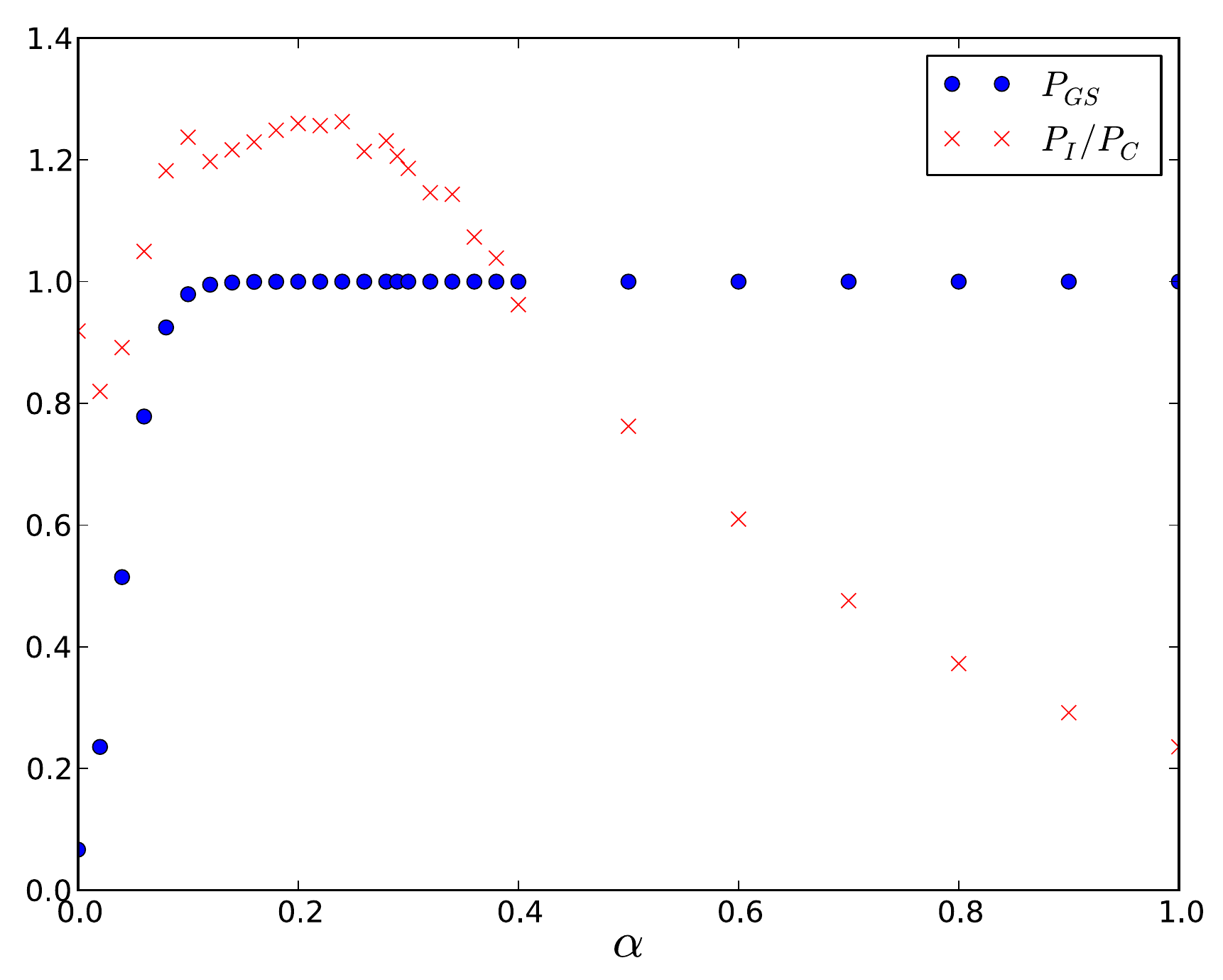}
\caption{Simulations results for the modified SSSV model. The model produces a signature similar to that of the D-Wave machine or quantum adiabatic master equation from Fig.~\ref{Vinci_results}. The model was simulated for 1,500 steps at the system temperature of $T=0.22\mbox{GHz}$. Ten thousand runs were performed for each value of $\alpha$.}\label{results}
\end{figure}
 
We also note that when $\alpha$ is small, the effects of systematic errors in the machine, such as imperfections in the calibration of the annealing schedule, will also become more dominant. Since the SSSV model does not attempt to model such systematic error, it is not surprising that it may fail to predict the machine's behavior in this regime. In fact, we are able to demonstrate that a simple modeling of systematic errors completely alters the SSSV model's behavior in this regime, so that it then reproduces the qualitative signature of the machine's behavior shown in \cite{Vinci}.

Fig.~\ref{results} shows the simulation results of the modified SSSV model in which there is a small independent Gaussian error in the calibration of the local field applied to each spin. To be more precise, the time-dependent Hamiltonian is defined as $H(t)=A(t)\sum_i \sin\theta_i - \sum_i (B(t)\cdot \alpha\cdot  h_i+\epsilon_i) \cos \theta_i- \sum_{i<j} B(t) \cdot \alpha\cdot  J_{ij} \cos \theta_i\cos \theta_j$ where $\epsilon_i \sim N(0,0.24)$.\footnote{We note that introducing similar Gaussian errors on the couplings does not seem to affect the simulation results.} We make no further attempt to improve the quantitative fit of these graphs (since detailed physical modeling of the machine is infeasible at the present time due to the limited access to the machine's internal mechanism), beyond noting that the set of examples in~\cite{Vinci} does not appear to provide a robust regime, in the sense described above, where the results of the D-Wave machine diverge from SSSV.

In a strict sense, establishing that a phenomenon is truly ``quantum" at a large scale is extremely challenging, since it involves ruling out all possible classical explanations. While this is not practically feasible, it is difficult to overemphasize the importance of carefully ruling out a range of classical models. Specifically, we hope that this note demonstrates the value of carefully considering elaborations of the rather rudimentary SSSV model while investigating how well it matches the behavior of a complex machine like D-Wave.

\begin{figure*}[!t]
\centering
\begin{subfigure}[b]{0.495\textwidth}
\includegraphics[width=\textwidth]{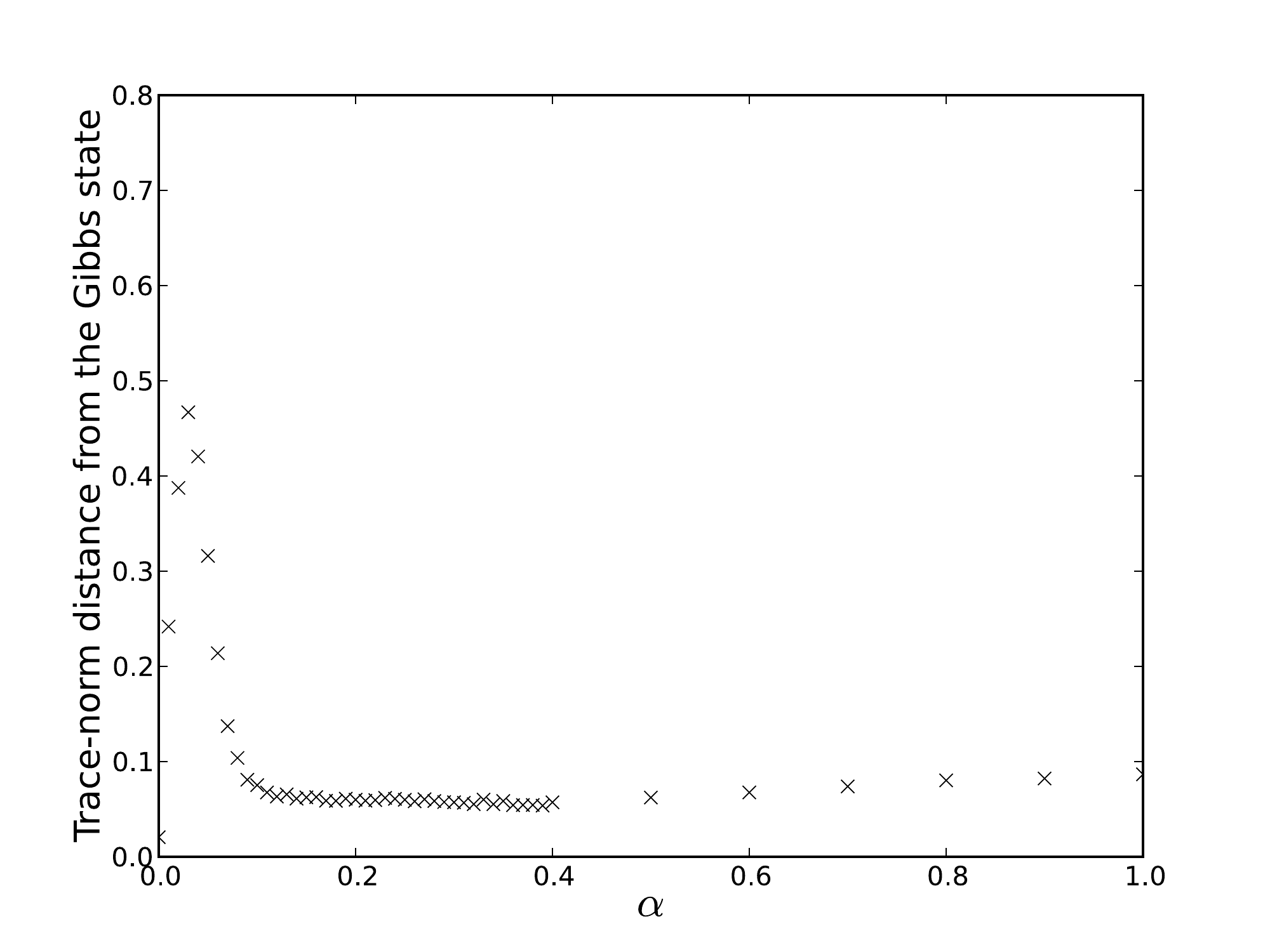}
\caption{Trace-norm distance from the Gibbs state.}\label{f1}
\end{subfigure}
\begin{subfigure}[b]{0.495\textwidth}
\includegraphics[width=\textwidth]{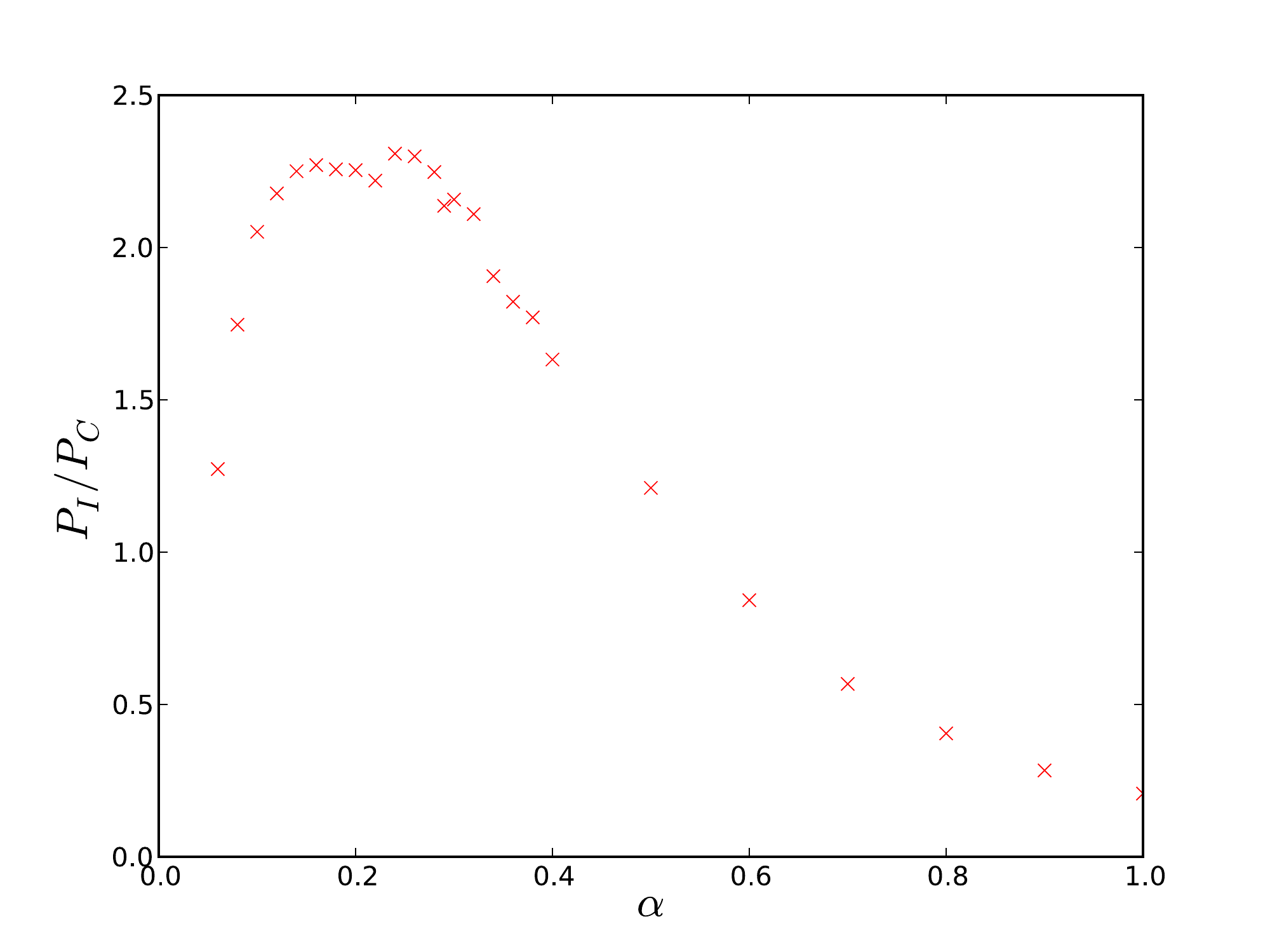}
\caption{$N=12$}\label{f2}
\end{subfigure}
\begin{subfigure}[b]{0.495\textwidth}
\includegraphics[width=\textwidth]{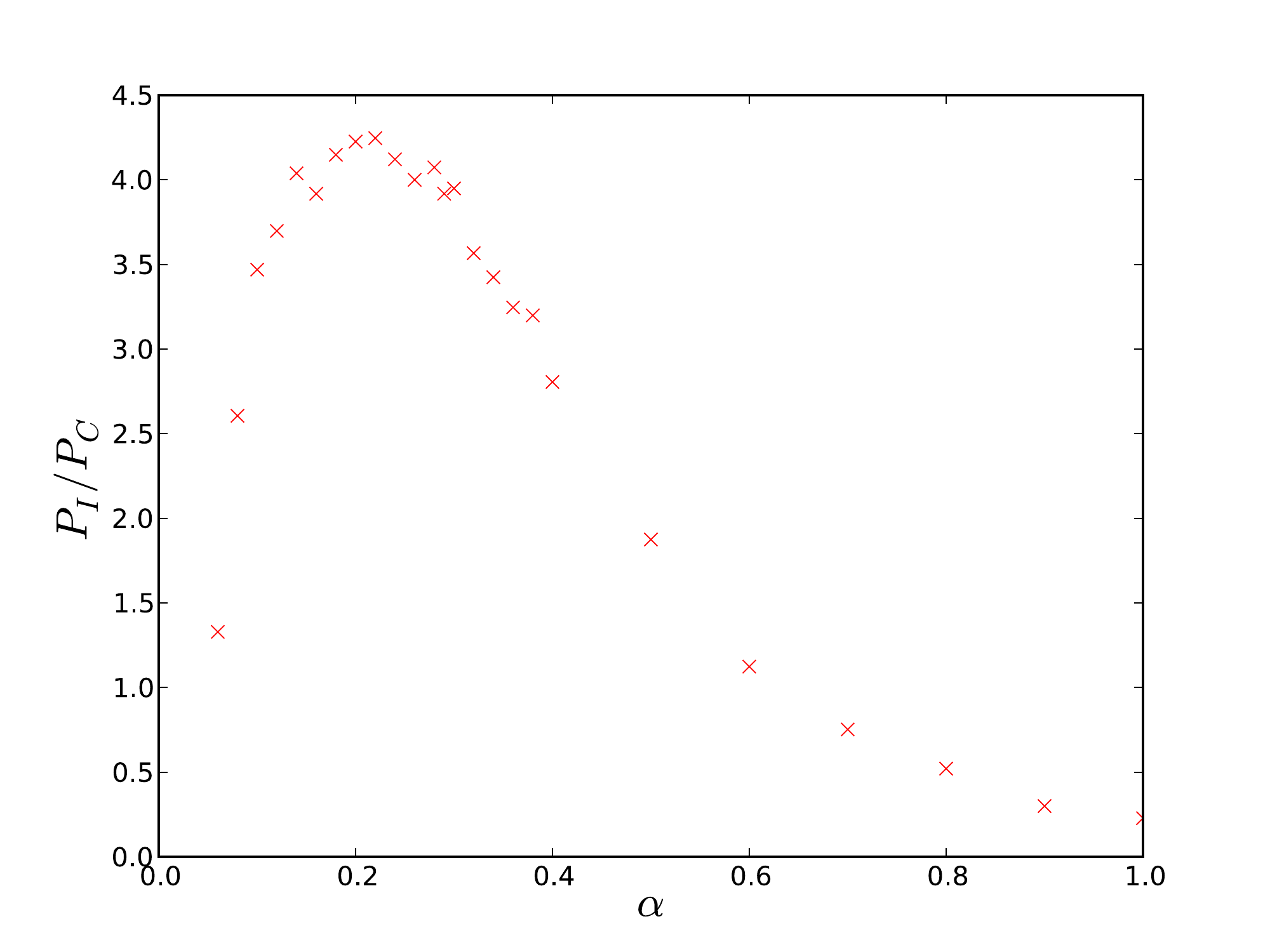}
\caption{$N=16$}\label{f3}
\end{subfigure}
\begin{subfigure}[b]{0.495\textwidth}
\includegraphics[width=\textwidth]{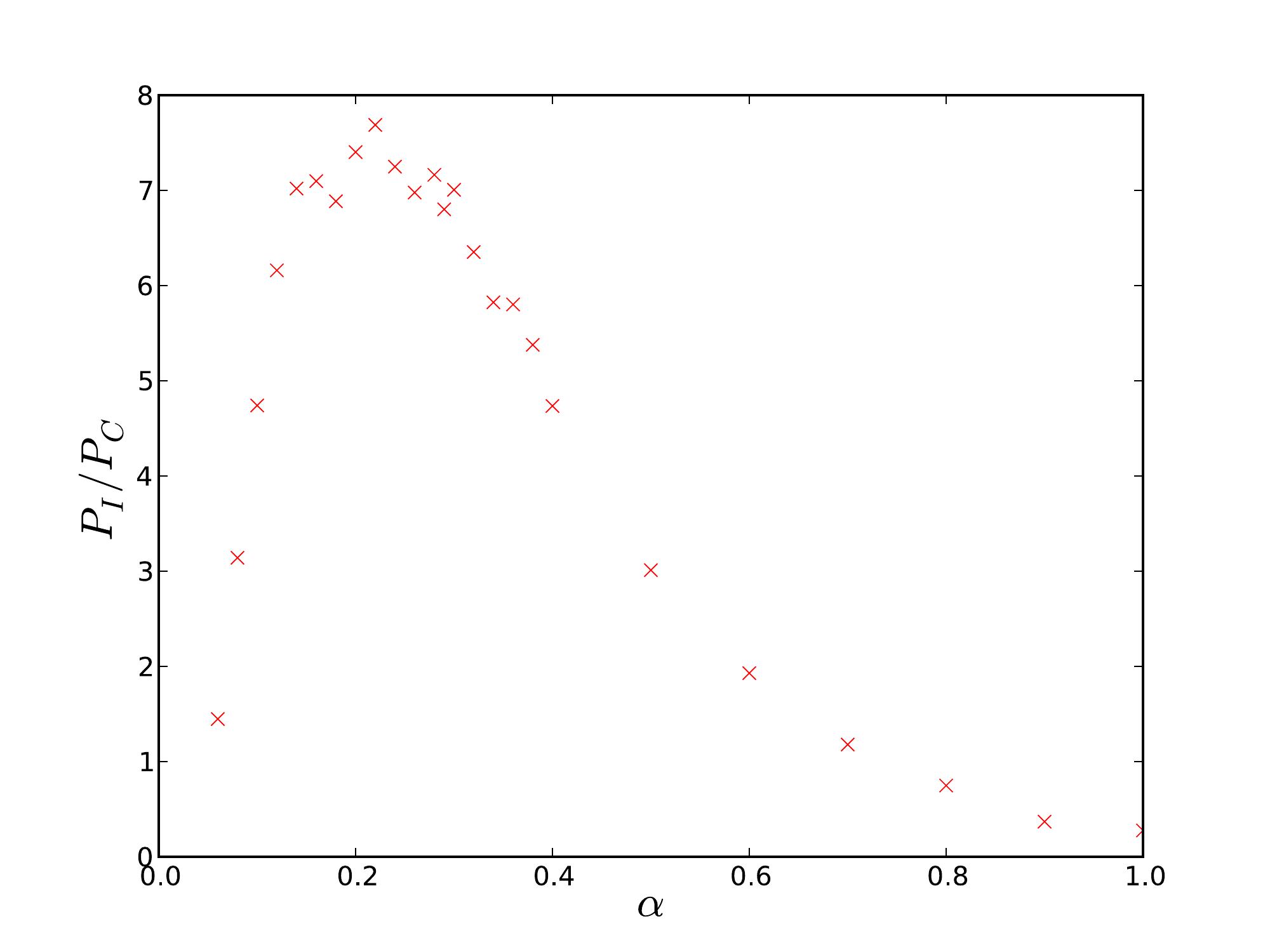}
\caption{$N=20$}\label{f4}
\end{subfigure}
\caption{Further simulations of the modified SSSV model reveal that it reproduces various other signatures suggested in \cite{Vinci}. For instance, Fig.~\ref{f1} exhibits a good qualitative resemblance with the experimental data presented in Fig.~14 of \cite{Vinci}. Figs.~\ref{f2}, \ref{f3}, and \ref{f4} show that the behavior demonstrated in Fig.~\ref{results} persists as the problem size scales up, which is consistent with the experimental results from Fig.~10 of \cite{Vinci}.}\label{additional}
\end{figure*}

\section*{Acknowledgments}
SWS and UV were supported by ARO Grant W911NF-09-1-0440 and NSF Grant CCF-0905626. 

\bibliographystyle{naturemag}
\bibliography{dwave}

\end{document}